\documentclass[useAMS,usenatbib]{mn2e}

\usepackage{graphics}
\usepackage{graphicx}
\usepackage{amsxtra}
\usepackage{array,longtable}
\usepackage{journals}
\usepackage{url}

\usepackage{latexsym}
\usepackage{amssymb}

\usepackage{ulem}
\usepackage{color}
\title
[Polar-ring galaxies]
{Polar-ring galaxies: the SDSS view on the symbiotic galaxies
}
\author[V.~Reshetnikov, F.~Combes]
{V.~Reshetnikov$^1$, F.~Combes$^2$ 
\\
$^1$St.Petersburg State University, Universitetskij pr. 28, 198504
St.Petersburg, Stary Peterhof, Russia \\
$^2$LERMA, Observatoire de Paris, CNRS (UMR 8112), 61, Av. de l'Observatoire, F-75014 Paris, France \\
}

\begin{document}

\date{Accepted 2014 December 5. Received 2014 November 23; in original form 2014 October 1}

\pagerange{\pageref{firstpage}--\pageref{lastpage}} \pubyear{2014}

\maketitle

\label{firstpage}

%%%%%%%%%%%%%%%%%%%%%%%%%%%%%%%%%%%%%%%%%%%%%%%%%%%%%%%%%%%%%%%%%%%%%%%
\begin{abstract}
%%%%%%%%%%%%%%%%%%%%%%%%%%%%%%%%%%%%%%%%%%%%%%%%%%%%%%%%%%%%%%%%%%%%%%%
Polar-ring galaxies are multi-spin systems, showing star formation
in a blue late-type component, perpendicular to a red early-type one, revealing how galaxy
formation can sometimes occur in successive steps. 
We perform two-dimensional decomposition in the $g$, $r$, $i$ bandpasses of 50 polar-ring
galaxies (PRGs) from the Sloan Digital Sky Survey. Each object was fit with
a S\'ersic host galaxy and a S\'ersic ring. Our general results are:
(i) The central (host) galaxies of the PRGs are non-dwarf
sub-$L^{\ast}$ galaxies with colors typical for early-type galaxies.
(ii) Polar structures in our sample are, on average,
fainter and bluer than their host galaxies.
(iii) In most galaxies, the stellar mass M$_*$ of the polar component is not negligible
in comparison with that of the host.
(iv) The distributions of the host galaxies on the size -- luminosity and Kormendy
diagrams are shifted by $\sim 1^m$ to fainter magnitudes in comparison
with E/S0 galaxies. It means that the PRGs hosts are more similar to quenched disks
than to ordinary early-type galaxies.
(v) All the PRGs in our sample are detected in mid-infrared by WISE, and we derive
from the 22$\mu$m luminosity their star formation rate (SFR). Their SFR/M$_*$ ratio
is larger than for the early-type galaxy sample of Atlas$^{\rm 3D}$, showing that the
star forming disk brings a significant contribution to the new stars.
Globally, PRGs appear frequently on the green valley in the
mass-color diagram, revealing the symbiotic character between a red-sequence host and
a blue cloud ring.
\end{abstract}


%%%%%%%%%%%%%%%%%%%%%%%%%%%%%%%%%%%%%%%%%%%%%%%%%%%%%%%%%%%%%%%%%%%%%%%
\section{Introduction}
%%%%%%%%%%%%%%%%%%%%%%%%%%%%%%%%%%%%%%%%%%%%%%%%%%%%%%%%%%%%%%%%%%%%%%%

Polar-ring galaxies (PRGs) are peculiar galaxies, consisting of a ring or disc of gas, stars and dust 
orbiting in a plane nearly perpendicular to the major axis of a central galaxy (\citealt{whit1990} = PRC). 
The central object is usually an early-type galaxy (ETG), poor in gas.
The polar ring is generally younger and looks similar to late-type galaxies, with large amount of HI
gas and young stars. PRGs are very rare objects -- the observed fraction of PRGs has been estimated 
to $\sim$0.5\% of all S0 galaxies (\citealt{whit1990, resh2011}).

To explain the structure of PRGs, several scenarios have been proposed: 
(1) the accretion scenario, where two interacting galaxies exchange mass (\citealt{reshsotn1997}); 
(2) the merging scenario (\citealt{bekki1997, bekki1998, bourcomb2003}); 
(3) the cosmic formation scenario, where the PRG form through the misaligned
accretion of gas from cosmic filaments (\citealt{maccio2006, brook2008}).   

PRGs are important laboratories to study the shape of their dark matter haloes (e.g. \citealt{comb2014}), and gain
insight into the galaxy formation through late infall, because of the high inclination of their off-plane 
structures and the evidence of very recent or occuring gas infall.

In summary, galaxies with polar rings are a unique class of extragalactic objects allowing to
investigate a wide range of problems, linked with the formation and evolution of galaxies, and to
study the properties of their dark haloes. The progress in the study of PRGs is strongly constrained
by a small number of known objects of this type.
In order to resolve this situation, a new catalogue of PRGs, supplementing the
catalogue of \citet{whit1990}, have been created (\citealt{moiseev2011} = SPRC). 
This catalogue consists of 275 candidates to polar-rings galaxies selected from the images of the 
Sloan Digital Sky Survey (SDSS DR7). 
The new catalogue significantly increases the number of genuine PRG candidates, and may serve as a 
good basis both for the further detailed study of individual galaxies, and for the statistical 
analysis of PRGs as a separate class of objects. 

The main purpose of this paper is to study the general photometric characteristics
of PRGs from the Sloan Survey. Throughout this article, we adopt a standard flat $\Lambda$CDM
cosmology with $\Omega_m$=0.3, $\Omega_{\Lambda}$=0.7, $H_0$=70 km\,s$^{-1}$\,Mpc$^{-1}$.

%%%%%%%%%%%%%%%%%%%%%%%%%%%%%%%%%%%%%%%%%%%%%%%%%%%%%%%%%%%%%%%%%%%%%%%
\section[]{The sample and the decomposition techniques}
\label{s_samples}
%%%%%%%%%%%%%%%%%%%%%%%%%%%%%%%%%%%%%%%%%%%%%%%%%%%%%%%%%%%%%%%%%%%%%%%

Our sample uncludes 46 objects from the ``best candidates'' of the SPRC. 
This group consists of 70 galaxies which are morphologically similar to well-studied ``classic'' 
PRGs from Whitmore et al. catalogue. These galaxies tend to have symmetrical
central body, resembling early-type galaxies (E/S0), and extended,
mostly visible ``edge-on'' structures crossing the central
galaxy at large angles. At the intersections of polar structures and
host galaxies are sometimes visible absorption band.
Initial analysis of the sample showed that at least two galaxies (SPRC-22 and
SPRC-43) are line-of-sight projections of two galaxies, and these objects were
excluded from further consideration. From the sample were excluded also too
weak and small by angular size objects and also galaxies, showing strongly asymmetric 
polar structures. The final sample was supplemented with 4 kinematically-confirmed 
PRGs from  \citet{whit1990} catalogue.  The list of the sample galaxies is presented in
the Table 1, where we provide the basic information about them. Currently, 12 of 50 galaxies
in the sample are kinematically-confirmed PRGs (8 from the SPRC -- see \citealt{moiseev2014}
and 4 from the PRC).

\begin{table}
\begin{minipage}{140mm}
\caption{Basic data for the PRGs sample}
\label{Table1}
\begin{tabular}{cllllll}
\hline 
\hline
 Galaxy & Redshift & r (mag) & $r_e~(\arcsec)$& $n$  & ring/host \\ 
(1) & (2)&(3)&(4)& (5) & (6) \\  \hline
SPRC-2  & --      & 14.37 & 6.9 & 2.5 & 0.31  \tabularnewline
SPRC-3$^*$  & 0.03683 & 15.93 & 2.8 & 2.1 & 0.90  \tabularnewline
SPRC-5  & 0.02786 & 17.09 & 1.5 & 1.8 & 0.19  \tabularnewline
SPRC-6  & 0.01867 & 17.02 & 2.0 & 1.6 & 0.11  \tabularnewline
SPRC-7$^*$  & 0.06007 & 16.42 & 0.9 & 2.3 & 1.3  \tabularnewline
SPRC-9  & 0.14473 & 17.47 & 2.3 & 1.5 & 0.36  \tabularnewline
SPRC-10$^*$ & 0.04244 & 16.26 & 2.9 & 1.2 & 0.11  \tabularnewline
SPRC-11 & 0.06580 & 15.67 & 3.7 & 3.7 & 0.11  \tabularnewline
SPRC-12 & 0.06247 & 17.13 & 1.9 & 2.2 & 0.49  \tabularnewline
SPRC-13 & 0.03177 & 15.23 & 3.1 & 3.0 & 0.39  \tabularnewline
SPRC-14$^*$ & 0.03188 & 14.70 & 4.4 & 4.5 & 0.24  \tabularnewline
SPRC-15 & 0.03427 & 14.46 & 3.3 & 3.0 & 0.22  \tabularnewline
SPRC-16 & 0.06012 & 16.98 & 2.4 & 1.5 & 0.32  \tabularnewline
SPRC-17 & 0.02640 & 14.88 & 7.0 & 3.2 & 0.05:  \tabularnewline
SPRC-18 & 0.08185 & 17.40 & 1.9 & 1.7 & 0.15  \tabularnewline
SPRC-19 & 0.10516 & 17.33 & 1.5 & 1.9 & 0.28  \tabularnewline
SPRC-20 & 0.07428 & 16.67 & 2.6 & 4.0 & 0.31  \tabularnewline
SPRC-23 & 0.02782 & 15.94 & 3.6 & 1.5 & 0.25  \tabularnewline
SPRC-24 & 0.04710 & 15.01 & 3.0 & 2.6 & 0.50  \tabularnewline
SPRC-25 & 0.07284 & 17.41 & 1.5 & 1.3 & 0.41  \tabularnewline
SPRC-27$^*$ & 0.04835 & 16.31 & 2.1 & 2.4 & 0.58  \tabularnewline
SPRC-28 & 0.07729 & 16.63 & 1.5 & 1.7 & 0.48  \tabularnewline
SPRC-29 & 0.04733 & 15.49 & 4.5 & 2.7 & 0.16  \tabularnewline
SPRC-30 & 0.07517 & 17.49 & 1.5 & 1.0 & 0.76  \tabularnewline
SPRC-31 & 0.04971 & 15.20 & 8.5 & 3.3 & 0.1:  \tabularnewline
SPRC-34 & 0.08131 & 16.97 & --  & --  & 0.37  \tabularnewline
SPRC-35 & 0.06757 & 17.16 & 1.9 & 0.8 & 0.98  \tabularnewline
SPRC-37 & 0.06762 & 16.11 & 8.3 & 4.4 & 0.20  \tabularnewline
SPRC-39$^*$ & 0.02933 & 15.99 & 4.5 & 3.1 & 0.27  \tabularnewline
SPRC-42 & 0.02337 & 15.13 & 4.0 & 2.7 & 1.3  \tabularnewline
SPRC-44 & 0.11344 & 16.55 & 2.9 & 1.5 & 0.20  \tabularnewline
SPRC-47 & 0.03123 & 14.63 & 8.5 & 1.9 & 0.15  \tabularnewline
SPRC-48 & 0.05630 & 15.32 & 7.9 & 2.5 & 0.12  \tabularnewline
SPRC-49 & 0.06838 & 15.91 & 2.6 & 1.5 & 0.05:  \tabularnewline
SPRC-51 & 0.07544 & 17.24 & 1.0 & 2.3 & 0.29  \tabularnewline
SPRC-53 & 0.08274 & 17.63 & 3.6 & 0.9 & 0.36  \tabularnewline
SPRC-55 & 0.08581 & 16.93 & 1.6 & 2.3 & 0.49  \tabularnewline
SPRC-56 & 0.05504 & 14.71 & 4.9 & 4.2 & 0.21  \tabularnewline
SPRC-57 & 0.07036 & 16.39 & 2.8 & 1.6 & 0.51  \tabularnewline
SPRC-58 & --      & 15.46 & 2.7 & 3.4 & 0.38  \tabularnewline
SPRC-59 & --      & 16.81 & 1.5 & 1.0 & 0.51  \tabularnewline
SPRC-63 & 0.07397 & 17.58 & --  & --  & 0.1:  \tabularnewline
SPRC-66 & 0.08738 & 16.10 & 2.0 & 1.7 & 0.33  \tabularnewline
SPRC-67$^*$ & 0.02777 & 14.07 & 5.1 & 3.5 & 0.46  \tabularnewline
SPRC-69$^*$ & 0.02469 & 15.24 & 2.1 & 1.8 & 0.28  \tabularnewline
SPRC-70 & 0.06870 & 17.59 & 2.4 & 0.8 & 1.3  \tabularnewline
PRC A-1$^*$   & 0.01835 & 14.57 & 2.6 & 3.1 & 0.38  \tabularnewline
PRC A-4$^*$   & 0.02341 & 14.41 & 6.7 & 5.1 & 0.14  \tabularnewline
PRC A-6$^*$   & 0.01798 & 14.76 & 3.7 & 2.5 & 0.49  \tabularnewline
PRC B-17$^*$  & 0.00426 & 13.80 & 11.9 & 1.6 & 0.19  \tabularnewline
\hline
\end{tabular}
\end{minipage}
   
\parbox[t]{73mm}{ Columns: \\
(1) name (SPRC or PRC), asterisks mark kinematically-confirmed galaxies, \\
(2) redshift (SDSS, NED), \\
(3) total r-band magnitude (present work),\\
(4) effective radius in the $r$ filter (present work), \\
(5) S\'ersic index in the $r$ band (present work), \\
(6) ring-to-central galaxy luminosity ratio in the r band (present work). }
 \end{table}

The 2D decomposition of the galaxy images from the sample 
of 50 PRGs was performed using the GALFIT code  (v3.0.4; \citealt{peng2010}).
GALFIT allows for various parametric functions (S\'ersic, Gaussian, etc.)
to be modelled simultaneously as either multiple subcomponents of a single
object, multiple objects in a frame or combination thereof. 

Galaxy images in the $g$, $r$ and $i$ bands were downloaded from the 
SDSS DR9 (\citealt{ahn2012}).  All stars, background and underground objects 
were masked. Nearby unsaturated stars were used as estimates of the PSF
at the objects positions. We fit a two-component model (host galaxy + ring)
to all galaxies in the sample. Both components were fitted by a single
S\'ersic function with seven free parameters (object centres, total magnitude,
effective radius $r_e$, S\'ersic index $n$, ellipticity, position angle).
In several cases (e.g. SPRC-7 -- galaxy with non edge-on ring -- see Fig.~1) we approximated
the ring by the inner-truncated model in order to describe clearly visible hole
in the ring component. Fig.~1 shows the results of decomposition for two galaxies
with different morphology (edge-on ring, non edge-on ring). 

The comparison of our derived apparent magnitudes of PRGs with the SDSS
{\it modelMag} values shows general agreement. Mean differences of the SDSS and
our magnitudes are +0$\fm$17$\pm$0$\fm$31 ($g$), +0$\fm$10$\pm$0$\fm$25 ($r$) and +0$\fm$09$\pm$0$\fm$23 ($i$).
After excluding 6 most deviated galaxies with magnitudes difference in the $g$ band
$\Delta g > +0\fm4$ (SPRC-3, 7, 20, 34, 37, 69), 
the differences reduce to +0$\fm$08$\pm$0$\fm$16 ($g$), +0$\fm$04$\pm$0$\fm$14 ($r$), and
+0$\fm$04$\pm$0$\fm$16 ($i$). 

For two galaxies with $\Delta g > +0\fm4$ we have found published results of surface photometry. 
\citet{brosch2010} have presented detailed study of SPRC-7 (SDSS J075234.33+292049.8). 
Our total magnitudes (16.87, 16.42, 16.08 in the $g$, $r$, $i$ passbands) are in good
agreement with Brosch et al. results (16.80, 16.41, 16.12). The parameters of the central galaxy
and the ring are in accord also. For the central galaxy \citet{brosch2010} give
such apparent magnitudes: 18.22, 17.35, 16.92 ($g$, $r$, $i$), the same values according to our
analysis are 18.28, 17.34, 16.90. The apparent magnitudes of the ring are 17.14, 17.00, 17.01
(\citealt{brosch2010}) and 17.21, 17.03, 16.78 (present work). \citet{fink2012} 
have published photometric data for 16 candidate PRGs, including SPRC-69 from our sample.
In the $r$ passband Finkelman et al. give such estimates of apparent magnitudes for SPRC-69: 
15.40 (host galaxy), 17.14 (ring), 15.20 (total). Our values are 15.51, 16.90 and 15.24
correspondingly. 

In the rest of the paper we will use the results of our photometry for all the sample PRGs
(see Table~1 for total magnitudes in the $r$ passband and some other characteristics). 

%%%%%%%%%%%%%%%%%%%%%%%%%%%%%%%%%%%%%%%%%%%%%%%%%%%%%%%%%%%%%%%%%%%%%%%
%%%% Fig 1
%%%%%%%%%%%%%%%%%%%%%%%%%%%%%%%%%%%%%%%%%%%%%%%%%%%%%%%%%%%%%%%%%%%%%%%
\begin{figure}
\includegraphics[width=8.3cm, angle=0, clip=]{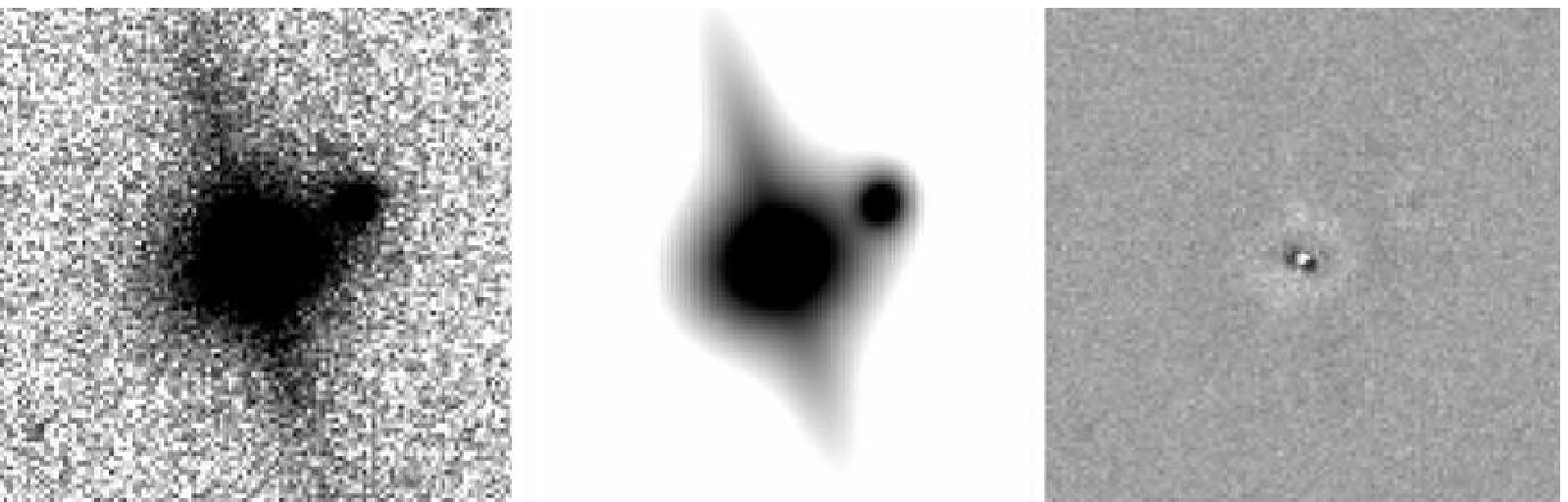}
\includegraphics[width=8.3cm, angle=0, clip=]{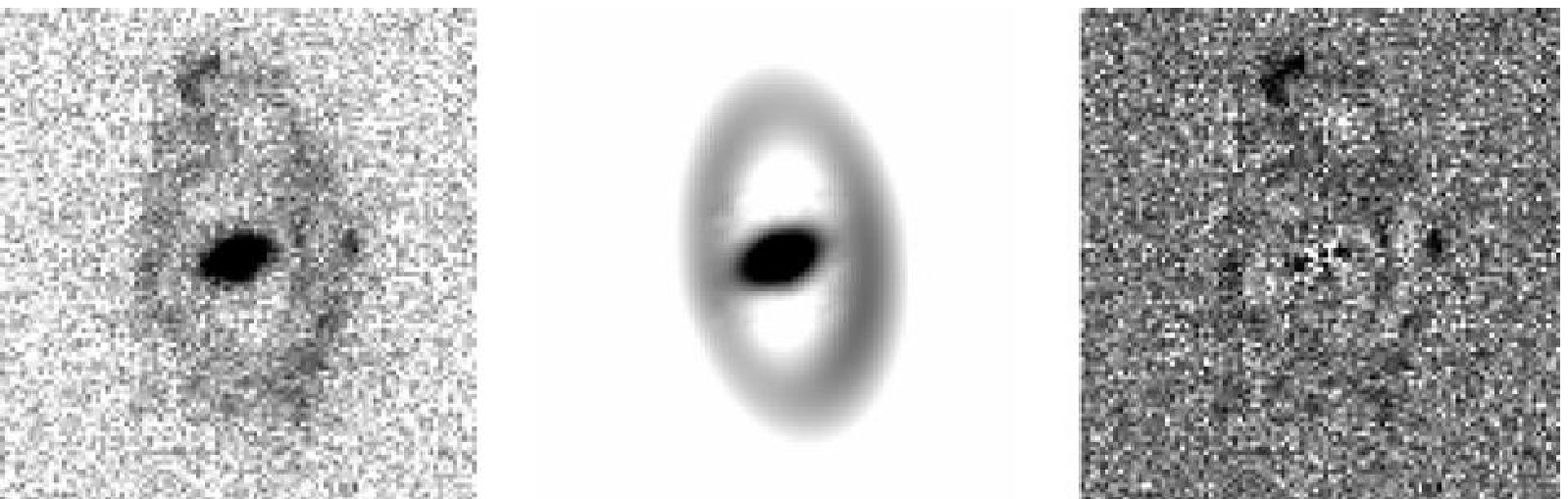}
\caption{Examples of the PRGs images modeling: SPRC-58 (top), SPRC-7 (bottom).
From left to right: original SDSS r-band image, 2D model, residual image.}
\end{figure}
%%%%%%%%%%%%%%%%%%%%%%%%%%%%%%%%%%%%%%%%%%%%%%%%%%%%%%%%%%%%%%%%%%%%%%% 

%%%%%%%%%%%%%%%%%%%%%%%%%%%%%%%%%%%%%%%%%%%%%%%%%%%%%%%%%%%%%%%%%%%%%%%
\section[]{Results}

\subsection{Central (host) galaxies}

Fig.~2a-c show distributions of central galaxies by absolute luminosity (corrected for 
Milky Way absorption according to \citealt{schfin2011} and $k$-correction from
\citealt{chil2010}), S\'ersic index and
physical effective radius. As one can see, they are non-dwarf sub-$L^{\ast}$ galaxies 
with average absolute magnitude $< M_r > = -20.34 \pm 1.04$.

%%%%%%%%%%%%%%%%%%%%%%%%%%%%%%%%%%%%%%%%%%%%%%%%%%%%%%%%%%%%%%%%%%%%%%%
%%%% Fig 2
%%%%%%%%%%%%%%%%%%%%%%%%%%%%%%%%%%%%%%%%%%%%%%%%%%%%%%%%%%%%%%%%%%%%%%%
\begin{figure}
\centering{
\includegraphics[width=12cm, angle=-90, clip=]{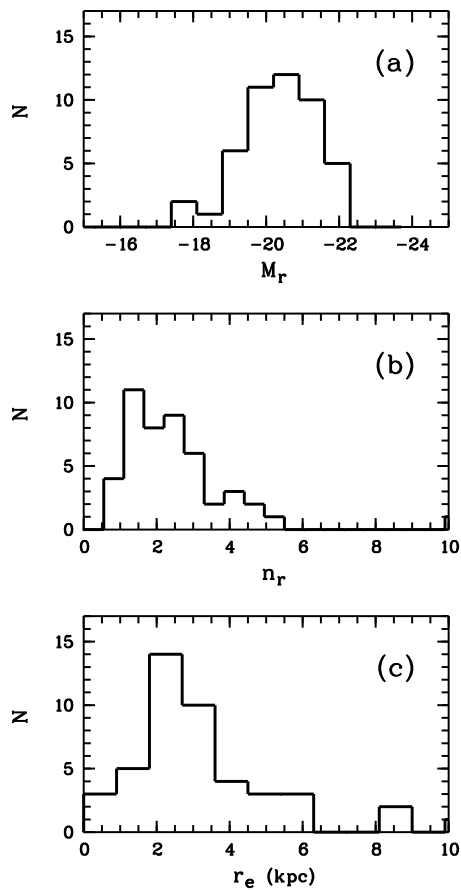}
\caption{Characteristics of PRGs hosts in the $r$ passband: (a) absolute magnitude,
(b) S\'ersic index, (c) effective radius.}
}
\end{figure}
%%%%%%%%%%%%%%%%%%%%%%%%%%%%%%%%%%%%%%%%%%%%%%%%%%%%%%%%%%%%%%%%%%%%%%% 

The distribution of S\'ersic indices of the host galaxies (Fig.~2b) indicates a broad range 
of morphologies, from disc-dominated ($n < 1.5$) to bulge-dominated ($n > 3$). 
It is important to note that $n$ is not a direct measure of the galaxy type and
it does not translate one to one to the bulge-to-total ratio. But the single S\'ersic
index is a reasonable statistical characteristic to separate late-type and early-type
galaxies (e.g. \citealt{bruce2012}).

The relative
fractions of disc-dominated and bulge-dominated brightness distributions are 16\%, 28\%, 
respectively, in the $r$ band and 26\%, 40\% ($i$ filter) in our sample of PRGs hosts.
These fractions are consistent with previous findings that late-type galaxies are
less frequent among PRGs in comparison with ETGs (e.g. \citealt{whit1990, whit1991, resh2011}).

Host galaxies of the PRGs show wide distribution of rest-frame optical colors 
(see the $g - r$ distribution in Fig.~3). In general, the $g - r$
color distribution looks similar to that presented by \citet{fink2012} 
(see their fig.~6) but the peak in our distribution is shifted to 
bluer color. (Most probably, this difference 
can be explained by the absence of $k$-correction in the \citealt{fink2012}
data.)

%%%%%%%%%%%%%%%%%%%%%%%%%%%%%%%%%%%%%%%%%%%%%%%%%%%%%%%%%%%%%%%%%%%%%%
%%%% Fig 3
%%%%%%%%%%%%%%%%%%%%%%%%%%%%%%%%%%%%%%%%%%%%%%%%%%%%%%%%%%%%%%%%%%%%%%%
\begin{figure}
\centering{
\includegraphics[width=5.0cm, angle=-90, clip=]{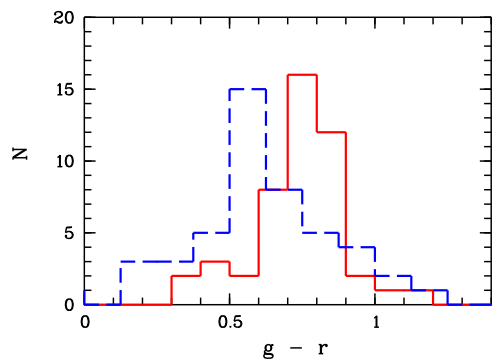}
\caption{Distribution of the sample galaxies over $g - r$ color of the host 
galaxies (red solid line) and of the rings (blue dashed line).}
}
\end{figure}
%%%%%%%%%%%%%%%%%%%%%%%%%%%%%%%%%%%%%%%%%%%%%%%%%%%%%%%%%%%%%%%%%%%%%%% 

The average colors of the PRGs hosts in our sample are $<g-r> = +0.74 \pm 0.16$, 
$<r-i> = +0.41 \pm 0.09$. Such colors correspond to the colors 
of S0 galaxies (\citealt{fukug2007}). \citet{fink2012} have compared
the $g - r$ color distribution of PRGs with that of in several control
samples of early-type galaxies (normal ETGs, blue ETGs, dusty ETGs). 
According to fig.~6 in their work, the average color of the PRGs hosts in our sample
matches the color of blue early-type galaxies better than typical colors of
other types of ETGs. It is important to note that optical colors of PRGs 
can be distorted by the inner absorption in the hosts and in the rings.
Near-infrared photometry of several PRGs was presented earlier 
by \citet{iodice2002a, iodice2002b}. They concluded that near-infrared colors
of host galaxies are bluer on average than those for standard early-type galaxies.

Fig.~4 shows standard scaling relations for the PRGs hosts. 
The size -- luminosity relation for the hosts is shown in Fig.~4a. PRGs are shifted
to lower luminosities in comparison with relation for early type galaxies (at a fixed
effective radius) and are located in the same region as spiral galaxies.
The characteristics of disc-dominated galaxies in the figure are taken from \citet{simard2011}.
\citet{simard2011} presented results of decomposition for more than 1 million galaxies
in the SDSS. We took their decomposition results with pure S\'ersic model and selected
large ($r_e \geq 1.\arcsec4$), nearby ($0.01 < z < 0.04$), disc-dominated ($n < 1.5$) 
galaxies for our Fig.~4.

Fig.~4b demonstrates the Kormendy relation (the mean surface brightness within effective radius
$r_e$ vs. $r_e$ in kpc) for PRGs and normal E/S0 galaxies. PRGs follow the standard relation
for E/S0 galaxies but, as in Fig.~4a, with notable shift (by $\sim1^m$) to lower 
surface brightnesses 
(at a fixed $r_e$). This shift is evident for the galaxies with various types of surface 
brightness distribution ($n < 3$ and $n > 3$). It can be concluded that PRG hosts are more
similar to quenched disks than to genuine elliptical galaxies.

%%%%%%%%%%%%%%%%%%%%%%%%%%%%%%%%%%%%%%%%%%%%%%%%%%%%%%%%%%%%%%%%%%%%%%%
%%%% Fig 4
%%%%%%%%%%%%%%%%%%%%%%%%%%%%%%%%%%%%%%%%%%%%%%%%%%%%%%%%%%%%%%%%%%%%%%%
\begin{figure}
\centering{
\includegraphics[width=11cm, angle=-90, clip=]{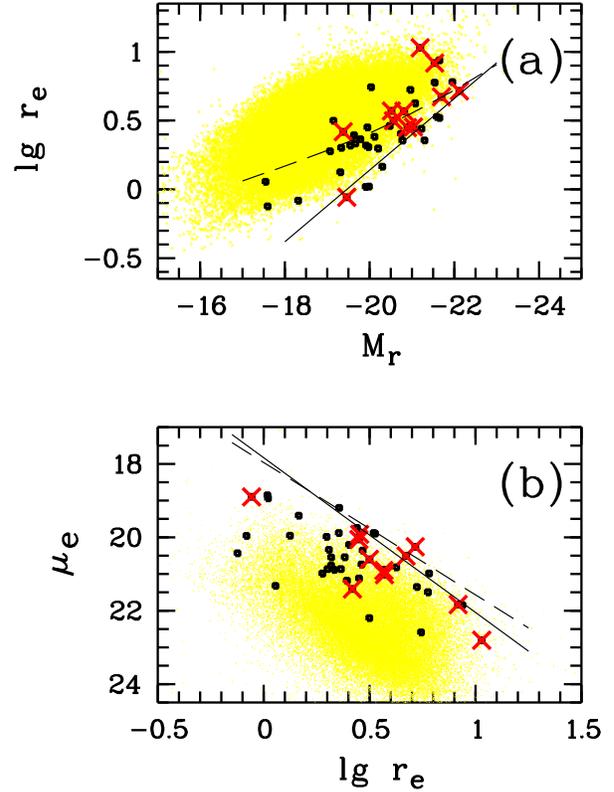}
\caption{Scaling relations for the host galaxies of PRGs in the $r$ band
(dots -- galaxies with S\'ersic index $n < 3$, red crosses -- with $n > 3$):
(a) absolute magnitude -- effective radius in kpc (solid line gives relation for
early type ($n > 2.5$) galaxies, dashed line -- for late type with $n < 2.5$, see
\citealt{shen2003}, yellow dots show the location of $\sim 40\,000$ disc-dominated galaxies
from \citealt{simard2011}), 
(b) Kormendy relation (solid line shows the relation for early type galaxies in Coma 
cluster (\citealt{hough2012}), dashed line represents mean relation for the SDSS early-type 
galaxies with absolute magnitudes $M_r$ between -18$\fm$5 and -22$\fm$0 according to 
\citealt{nn2008}, yellow dots are as above).}
}
\end{figure}
%%%%%%%%%%%%%%%%%%%%%%%%%%%%%%%%%%%%%%%%%%%%%%%%%%%%%%%%%%%%%%%%%%%%%%% 

\subsection{Polar structures}

Polar structures in our sample of PRGs are, on average, fainter ($< M_r > = -18.90 \pm 1.28$)
and bluer ($<g-r> = +0.61 \pm 0.25$, $<r-i> = +0.33 \pm 0.22$) than their host galaxies
(Fig.~3; see also \citealt{resh1994, iodice2002b, fink2012}).
Taking into account internal extinction in the polar rings (most of
them are seen almost edge-on), the difference in colors must be larger. Observed colors
of rings are usual for normal spiral galaxies (e.g. \citealt{fukug2007}).

As it was shown earlier by \citet{fink2012}, observed luminosities of
the hosts and the rings are correlated -- more luminous host galaxies possess
more luminous rings, on average (Fig.~5). There is also a weak mutual (ring vs. host) 
correlation of colors but much weaker than in \citet{fink2012}. The other possible trend
is between the relative size of ring ($D_{ring}/D_{host}$ -- diameter of ring normalized 
by diameter of host galaxy) and absolute magnitude of host galaxy -- see Fig.~6.
(The rings sizes were taken from \citealt{smirmois2013}). Among PRGs with
bright central galaxies we see relatively compact rings, while fainter central
objects are surrounded by relatively small as far as more extended rings:
for bright hosts with $M_r \leq -21$ the mean relative size of rings is
1.4$\pm$0.4, for faint ones with $M_r \geq -20$ the mean value is 2.3$\pm$1.2.

%%%%%%%%%%%%%%%%%%%%%%%%%%%%%%%%%%%%%%%%%%%%%%%%%%%%%%%%%%%%%%%%%%%%%%
%%%% Fig 5
%%%%%%%%%%%%%%%%%%%%%%%%%%%%%%%%%%%%%%%%%%%%%%%%%%%%%%%%%%%%%%%%%%%%%%%
\begin{figure}
\centering{
\includegraphics[width=5.5cm, angle=-90, clip=]{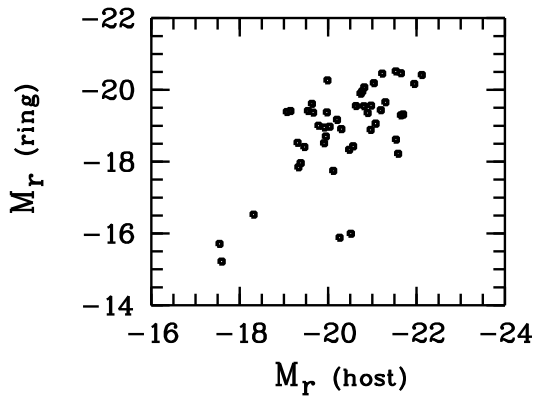}
\caption{Absolute $r$-band magnitudes of host galaxies versus rings.}
}
\end{figure}
%%%%%%%%%%%%%%%%%%%%%%%%%%%%%%%%%%%%%%%%%%%%%%%%%%%%%%%%%%%%%%%%%%%%%%% 

We estimated the stellar masses of the rings and the host galaxies following the
\citet{bell2003} approach, which combines a galaxy's luminosity with a mass-to-light
ratio from a color measurement. Our calculations employ the $g - r$ colors and $r$-band 
luminosities. The results are shown in Fig.~7. As one can see, stellar mass of
a polar component is not negligible in comparison with mass of a host in most
galaxies. In some galaxies the masses of two components are comparable. 

%%%%%%%%%%%%%%%%%%%%%%%%%%%%%%%%%%%%%%%%%%%%%%%%%%%%%%%%%%%%%%%%%%%%%%
%%%% Fig 6
%%%%%%%%%%%%%%%%%%%%%%%%%%%%%%%%%%%%%%%%%%%%%%%%%%%%%%%%%%%%%%%%%%%%%%%
\begin{figure}
\centering{
\includegraphics[width=5.5cm, angle=-90, clip=]{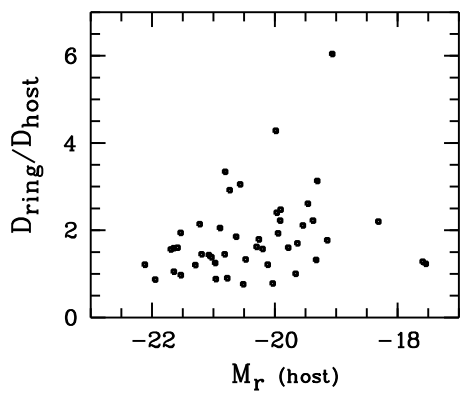}
\caption{Absolute $r$-band magnitudes of hosts versus relative sizes of polar rings.}
}
\end{figure}
%%%%%%%%%%%%%%%%%%%%%%%%%%%%%%%%%%%%%%%%%%%%%%%%%%%%%%%%%%%%%%%%%%%%%%% 

%%%%%%%%%%%%%%%%%%%%%%%%%%%%%%%%%%%%%%%%%%%%%%%%%%%%%%%%%%%%%%%%%%%%%%
%%%% Fig 7
%%%%%%%%%%%%%%%%%%%%%%%%%%%%%%%%%%%%%%%%%%%%%%%%%%%%%%%%%%%%%%%%%%%%%%%
\begin{figure}
\centering{
\includegraphics[width=5.5cm, angle=-90, clip=]{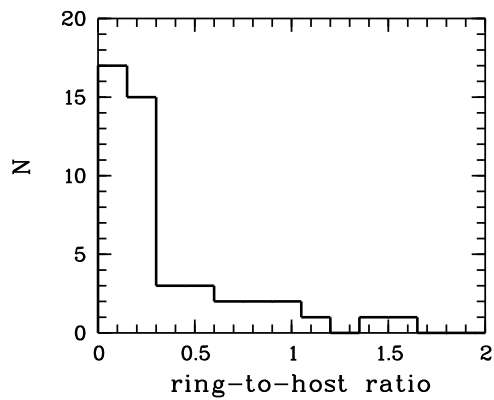}
\caption{Distribution of the sample galaxies over the ratio of stellar mass of
a ring to stellar mass of a host galaxy.}
}
\end{figure}
%%%%%%%%%%%%%%%%%%%%%%%%%%%%%%%%%%%%%%%%%%%%%%%%%%%%%%%%%%%%%%%%%%%%%%% 

Fig.~8 presents the observed distribution of angle between the ring 
and the central galaxy. We see that rings in most galaxies are within 20$\degr$ from
polar orientation, so our objects are indeed ``polar''-ring galaxies. Relative
fraction of such galaxies in the combined sample (our work + \citealt{whit1991}) 
is 75$\%$ (55 of 73 galaxies). Fig.~8 shows the {\it projected, apparent}
angles between the host galaxy and the ring. Recently, \citet{smirmois2013}
performed analysis of {\it spatial} angles between two components in the sample of 78
galaxies from the SPRC and PRC. They came to the same conclusion -- in the majority
of PRGs the outer structures lie in the plane close to polar within 10$\degr$--20$\degr$.

%%%%%%%%%%%%%%%%%%%%%%%%%%%%%%%%%%%%%%%%%%%%%%%%%%%%%%%%%%%%%%%%%%%%%%
%%%% Fig 8
%%%%%%%%%%%%%%%%%%%%%%%%%%%%%%%%%%%%%%%%%%%%%%%%%%%%%%%%%%%%%%%%%%%%%%%
\begin{figure}
\centering{
\includegraphics[width=5.5cm, angle=-90, clip=]{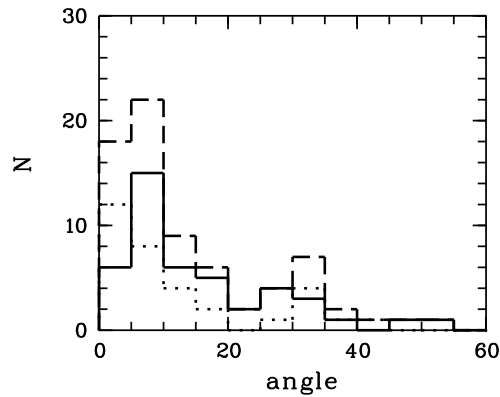}
\caption{Angular distance away from perpendicularity of the host galaxy and
the ring. Solid line -- our measurements ($g$ passband), dotted line -- the data from 
Whitmore (1991), dashed line -- combined sample.}
}
\end{figure}
%%%%%%%%%%%%%%%%%%%%%%%%%%%%%%%%%%%%%%%%%%%%%%%%%%%%%%%%%%%%%%%%%%%%%%% 

\subsection{Mid-infrared luminosity and star formation rate}

All galaxies of our sample were detected in the four bands
(3.4, 4.6, 12 and 22$\mu$m) of WISE, and we derived the  22$\mu$m
luminosity in Table 2. From this luminosity, an indication of the star formation rate (SFR)
has been derived, following the calibration of \citet{calz2007}, also displayed
in Table 2. The SFR for PRGs are distributed on the high side of what
is found for typical early-type galaxy samples. In Fig.~9, we compare the
distribution of the ratio between  22$\mu$m and K-band luminosities,
with the early-type sample of Atlas$^{\rm 3D}$ (\citealt{davis2014}).
It is clear that the PRG galaxies form relatively more stars than normal
early-type galaxies. This shows that the presence of the polar disks,
although faint, are significative in this respect.

%%%%%%%%%%%%%%%%%%%%%%%%%%%%%%%%%%%%%%%%%%%%%%%%%%%%%%%%%%%%%%%%%%%%%%
%%%% Fig 9
%%%%%%%%%%%%%%%%%%%%%%%%%%%%%%%%%%%%%%%%%%%%%%%%%%%%%%%%%%%%%%%%%%%%%%%
\begin{figure}
\centering{
\includegraphics[width=6.0cm, angle=-90, clip=]{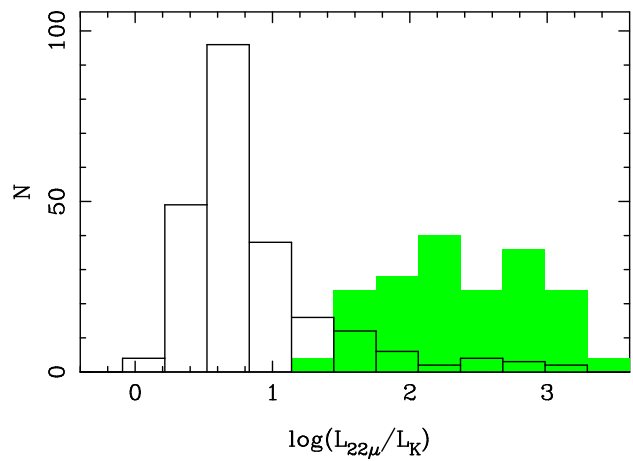}
\caption{Distribution of our PRG sample galaxies (green) compared
to the Atlas$^{\rm 3D}$ sample (black line), as a function of the
ratio between 22$\mu$m and K-band luminosities. The numbers for the PRGs
have been multiplied by 4 to compare with the much larger Atlas$^{\rm 3D}$ sample.}
}
\end{figure}
%%%%%%%%%%%%%%%%%%%%%%%%%%%%%%%%%%%%%%%%%%%%%%%%%%%%%%%%%%%%%%%%%%%%%%% 

Also, Fig.~10 shows the distribution of polar rings and hosts in the color-mass diagram.
Objects with SFR higher than 2.5 M$_\odot$/yr are plotted with large symbols,
PRG rings are clearly in the blue cloud and the hosts are in the yellow one. 
In Fig.~11, the global colors are plotted in the same
diagram (for polar rings and hosts together). The objects at high SFR are
distributed in the green valley mainly, showing that this location is due to the
symbiose between two different sub-systems.

%%%%%%%%%%%%%%%%%%%%%%%%%%%%%%%%%%%%%%%%%%%%%%%%%%%%%%%%%%%%%%%%%%%%%%%
%%%% Fig 10
%%%%%%%%%%%%%%%%%%%%%%%%%%%%%%%%%%%%%%%%%%%%%%%%%%%%%%%%%%%%%%%%%%%%%%%
\begin{figure}
\includegraphics[width=7.0cm, angle=-90, clip=]{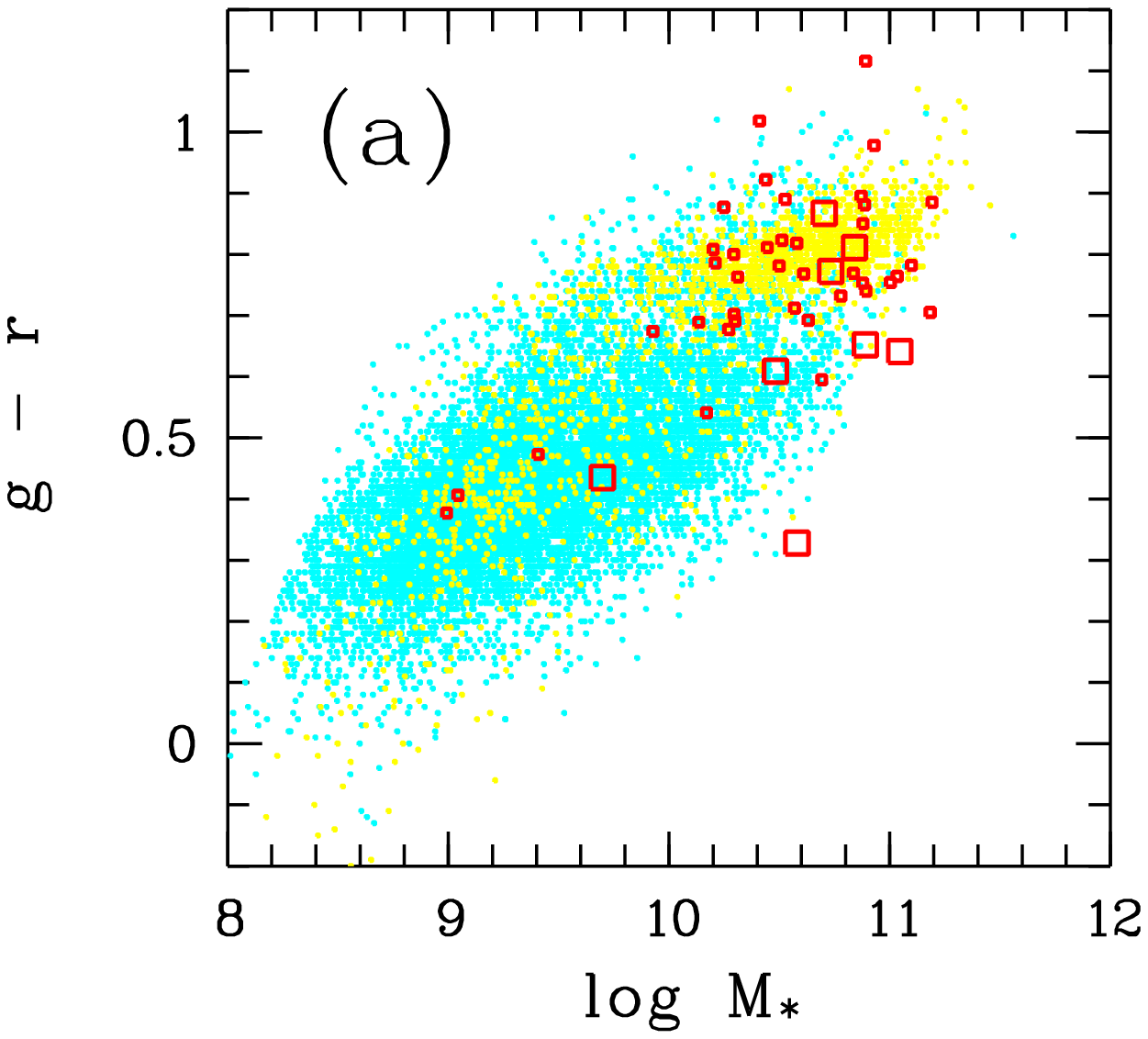}
\includegraphics[width=7.0cm, angle=-90, clip=]{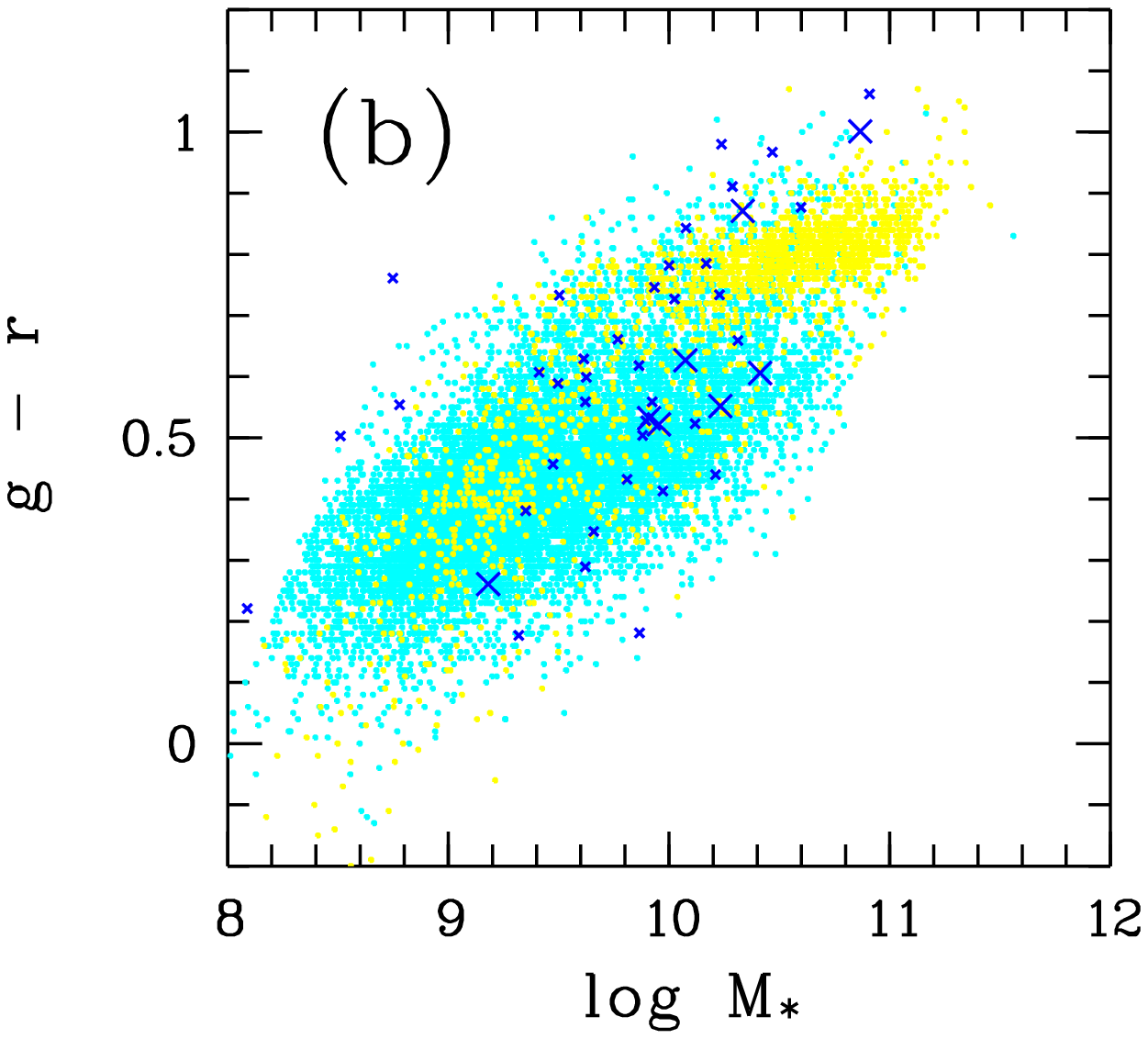}
\caption{The $g - r$ color -- stellar mass (in units of M$_\odot$) diagram (a) of the host galaxies
(red symbols) and (b) of the polar rings (dark blue symbols). Galaxies with global star-formation
rate $>$2.5 M$_\odot$/yr are marked by large symbols. Blue and yellow points show
locations of nearby ($z = 0.01 - 0.03$) late-type (with S\'ersic index n=0.8-1.2) and early-type
(n=3.5-4.5) galaxies from \citet{simard2011} correspondingly. Galaxies with absolute magnitudes
$-16 \ge  M_r \ge -23$ are shown only.}
\end{figure}
%%%%%%%%%%%%%%%%%%%%%%%%%%%%%%%%%%%%%%%%%%%%%%%%%%%%%%%%%%%%%%%%%%%%%%% 

%%%%%%%%%%%%%%%%%%%%%%%%%%%%%%%%%%%%%%%%%%%%%%%%%%%%%%%%%%%%%%%%%%%%%%
%%%% Fig 11
%%%%%%%%%%%%%%%%%%%%%%%%%%%%%%%%%%%%%%%%%%%%%%%%%%%%%%%%%%%%%%%%%%%%%%%
\begin{figure}
\centering{
\includegraphics[width=6.0cm, angle=-90, clip=]{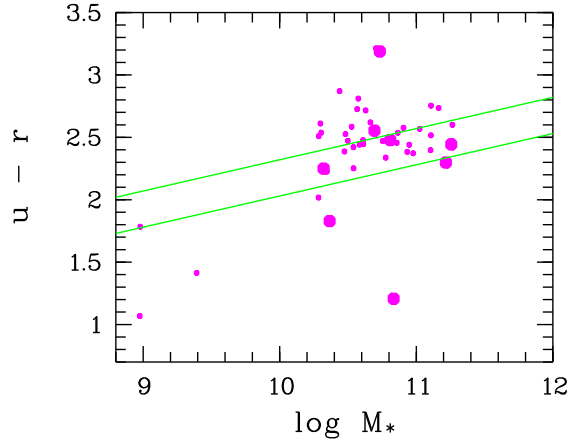}
\caption{The $u - r$ -- mass diagram of the PRGs according to the SDSS total magnitudes
corrected for the Milky Way absorption and $k$-correction. 
Large symbols show galaxies with star-formation rate $>$2.5 M$_\odot$/yr,
green lines show the green valley (\citealt{schaw2014}). }
}
\end{figure}
%%%%%%%%%%%%%%%%%%%%%%%%%%%%%%%%%%%%%%%%%%%%%%%%%%%%%%%%%%%%%%%%%%%%%%% 

\section{Discussion and conclusions}

We have performed a survey of photometric characteristics in a sample of 50
PRGs selected from the SDSS. In general, we have confirmed previous results but
on the basis of a much larger and uniform sample. 

We have found that
central objects of PRGs look like early-type galaxies by morphology and optical colors.
Polar structures demonstrate optical colors usual for spiral galaxies. Typical stellar
masses of the rings in our sample are about 10$^{9}$--10$^{10}$\,M$_{\odot}$ and 
they contribute significantly to the total stellar mass of PRGs. 

The most interesting results can be seen in Fig.~4. Fig.~4b shows the Kormendy relation
for the PRG hosts. It is evident that the location of PRGs is shifted from the relation
for E/S0 galaxies. This feature was first noted by \citet{resh1994} for bulges of 6 
PRGs in the $B$ passband. They discussed this point and proposed that 
the rings projection distorts the surface brightness distribution of the central regions
of galaxies. Projection of an absorbing ring can reduce the observed
luminosity of a galaxy and shift its characteristics on the Kormendy relation. 
\citet{iodice2002b} found a similar behavior in the $K$ passband for the sample 
of 5 PRGs. But, since Iodice et al. used observations in the infrared where the 
dust absorption is minimal, they concluded that this shift can be explained by the
small size of the PRG bulges in comparison with normal early-type galaxies.  
From the other side, our Fig.~4a demonstrates that the PRG hosts are not
compact -- they are more extended in comparison with E/S0 galaxies of the same
luminosities and are located in the locus of spirals. This is true not only of late-type 
galaxies but also of early-type,
bulge-dominated galaxies with $n > 3$. Therefore, an alternative explanation 
for the shift of the PRGs on the Kormendy relation is that their hosts are
fainter (in surface brightness and in luminosity) in comparison with normal
early-type galaxies of the same size. 

\begin{table}
%\begin{minipage}{140mm}
\caption{WISE and 2MASS data for our PRG sample}
\label{Table2}
\begin{tabular}{cllll}
\hline 
\hline
 Galaxy & F$_{22}$ & L$_{22}$ & SFR & L$_K$ \\ 
(1) & (2)&(3)&(4)& (5) \\  \hline
SPRC-3  &  14.41 &  42.792 &  0.94 & 10.32\\
SPRC-5  &   9.13 &  42.345 &  0.38 &  9.36\\
SPRC-6  &   6.74 &  41.859 &  0.14 &  9.05\\
SPRC-7  &   7.71 &  42.961 &  1.33 & 10.32\\
SPRC-9  &  11.58 &  43.954 & 10.1 & 11.28\\
SPRC-10 &  10.89 &  42.797 &  0.95 & 10.63\\
SPRC-11 &  10.46 &  43.176 &  2.06 & 11.12\\
SPRC-12 &  13.51 &  43.240 &  2.35 & 10.27\\
SPRC-13 &   8.04 &  42.406 &  0.43 & 10.48\\
SPRC-14 &  10.32 &  42.518 &  0.54 & 10.83\\
SPRC-15 &   6.42 &  42.376 &  0.40 & 11.02\\
SPRC-16 &  11.58 &  43.138 &  1.91 & 10.62\\
SPRC-17 &  10.47 &  42.357 &  0.39 & 10.64\\
SPRC-18 &  18.05 &  43.613 &  5.03 & 10.59\\
SPRC-19 &  10.06 &  43.592 &  4.81 & 10.76\\
SPRC-20 &  13.92 &  43.411 &  3.33 & 10.84\\
SPRC-23 &  61.94 &  43.175 &  2.06 & 10.64\\
SPRC-24 &  12.07 &  42.935 &  1.26 & 11.14\\
SPRC-25 &   6.78 &  43.081 &  1.70 & 10.58\\
SPRC-27 &   6.99 &  42.722 &  0.82 & 10.53\\
SPRC-28 &   6.43 &  43.112 &  1.81 & 10.67\\
SPRC-29 &   5.55 &  42.602 &  0.64 & 11.05\\
SPRC-30 &   8.23 &  43.193 &  2.14 & 10.54\\
SPRC-31 &  13.00 &  43.016 &  1.49 & 10.97\\
SPRC-34 &  11.43 &  43.408 &  3.31 & 10.47\\
SPRC-35 &   7.39 &  43.049 &  1.60 & 10.38\\
SPRC-37 &  10.39 &  43.198 &  2.16 & 11.00\\
SPRC-39 &  13.02 &  42.545 &  0.57 & 10.27\\
SPRC-42 &   8.25 &  42.145 &  0.25 & 10.39\\
SPRC-44 &   8.71 &  43.600 &  4.90 & 11.47\\
SPRC-47 &  10.32 &  42.500 &  0.52 & 11.00\\
SPRC-48 &   8.08 &  42.922 &  1.23 & 11.13\\
SPRC-49 &   8.39 &  43.116 &  1.83 & 11.23\\
SPRC-51 &   4.76 &  42.959 &  1.33 & 10.81\\
SPRC-53 &   5.79 &  43.129 &  1.88 & --   \\
SPRC-55 &   4.97 &  43.097 &  1.76 & 10.82\\
SPRC-56 &   4.91 &  42.685 &  0.76 & 11.38\\
SPRC-57 &   6.44 &  43.027 &  1.52 & 10.82\\
SPRC-63 &   6.61 &  43.084 &  1.71 & 10.55\\
SPRC-66 & 117.91 &  44.488 & 29.9 & 11.29\\
SPRC-67 &   4.00 &  41.984 &  0.18 & 10.90\\
SPRC-69 &  14.93 &  42.451 &  0.47 & 10.43\\
SPRC-70 &  13.82 &  43.336 &  2.86 & 10.37\\
PRCA-1  &   7.61 &  41.896 &  0.15 & 10.40\\
PRCA-4  &  10.86 &  42.266 &  0.32 & 10.67\\
PRCA-6  &   5.83 &  41.763 &  0.12 & 10.40\\
PRCB-17 &  21.92 &  41.077 &  0.03 &  9.02\\
\hline
\end{tabular}
%\end{minipage}
   
\parbox[t]{73mm}{ Columns: \\
(1) name (SPRC or PRC), \\
(2) Flux (22 $\mu$m) in mJy, \\
(3) log Luminosity (22 $\mu$m) in erg/s,\\
(4) SFR in M$_\odot$/yr, \\
(5) log L$_K$ (L$_\odot$).  }
 \end{table} 
 
In general, as one can derive from Fig.~4, PRG hosts may represent a transitional class 
between early- and late-type galaxies. In this sence, they look similar to 
blue-sequence E/S0 galaxies described by \citet{kann2009}. Blue-sequence E/S0
galaxies are shifted from the stellar mass -- radius relation for normal ETGs and
are located between late-type and early-type galaxies in this plane (see fig.~9
in \citealt{kann2009}). At a fixed stellar mass, blue-sequence E/S0 galaxies 
are somewhat larger in size than E/S0 galaxies. This behavior is similar for
PRGs: their hosts are larger at a fixed luminosity and they are located between
spirals and ellipticals on the Kormendy relation (Fig.~4a,b). 

\citet{kann2009} have noted that blue-sequence E/S0s are often associated with
counter-rotating gas and polar rings. We confirmed this conclusion from the opposite
point of view -- central galaxies of PRGs look similar to blue-sequence E/S0s.

One interesting result is also the distribution of the star formation rate for PRGs.
The ratio  between  22$\mu$m and K-band luminosities is significantly larger
for our PRGs sample than for the early-type sample of Atlas$^{\rm 3D}$.
PRG galaxies are a symbiotic objects, with two different sub-systems living together,
one in the red sequence, the other in the blue cloud, and the global ensemble
appears to be in the green valley.

Our results show that the structure of the PRGs hosts are more
similar to quenched disks than to 
ordinary early-type galaxies. More detailed observations and sophisticated 
modeling required to solve the puzzles of this mysterious class of extragalactic objects.

%%%%%%%%%%%%%%%%%%%%%%%%%%%%%%%%%%%%%%%%%%%%%%%%%%%%%%%%%%%%%%%%%%%%%%%%
\section*{Acknowledgments}
%%%%%%%%%%%%%%%%%%%%%%%%%%%%%%%%%%%%%%%%%%%%%%%%%%%%%%%%%%%%%%%%%%%%%%%%

We thank the referee for useful comments.
VR express gratitude for the grant of the Russian 
Foundation for Basic Researches number 13-02-00416. This work was partly supported by 
St. Petersburg State University research grants 6.0.160.2010, 6.0.163.2010, and 6.38.71.2012. 
VR acknowledges the hospitality of Paris Observatory, where a large part of this work was done.  
FC acknowledges the European Research Council
for the Advanced Grant Program Number 267399-Momentum.
This publication makes use of data products from the Wide-field Infrared Survey Explorer, 
which is a joint project of the University of California, Los Angeles, and the Jet
Propulsion Laboratory/California Institute of Technology, funded
by the National Aeronautics and Space Administration, and data
products from the Two Micron All Sky Survey, which is a joint
project of the University of Massachusetts and the Infrared
Processing and Analysis Center/California Institute of Technology,
funded by the National Aeronautics and Space Administration and the 
National Science Foundation.

\bibliographystyle{mn2e}
\bibliography{art}

\label{lastpage}

\end{document}